# Enhanced heat dissipation and lowered power consumption in electronics using two-dimensional hexagonal boron nitride coatings

*Karthik R, Ashutosh Srivastava, Soumen Midya, Akbar Shanu, Surbhi Slathia, Sajith Vandana, Punathil Raman Sreeram\*, Swastik Kar\*, Nicholas R. Glavin, Ajit K Roy, Abhishek Kumar Singh\*, Chandra Sekhar Tiwary\**

**Karthik R, Punathil Raman Sreeram, Chandra Sekhar Tiwary**

*Department of Metallurgical and Materials Engineering, Indian Institute of Technology Kharagpur, West Bengal 721302, India.*

E-mail: sreerampunam@metal.iitkgp.ac.in (P. R. Sreeram),

chandra.tiwary@metal.iitkgp.ac.in (C. S. Tiwary)

**Ashutosh Srivastava, Soumen Midya, Abhishek Kumar Singh**

*Materials Research Centre, Indian Institute of Science, Bangalore 560012, India*

E-mail: *abhishek@iisc.ac.in* (AKS)

**Akbar Shanu, Sajith Vandana**

*Department of Materials Science and Engineering, National Institute of Technology Calicut, Kerala 673601, India*

**Surbhi Slathia, Chandra Sekhar Tiwary**

*School of Nano Science and Technology, Indian Institute of Kharagpur, West Bengal, 721302, India*

**Swastik Kar**

*Department of Physics, Quantum Materials and Sensing Institute, Department of Chemical Engineering, Northeastern University, Boston, Massachusetts 02115, USA*

E-mail: *s.kar@northeastern.edu*

**Nicholas R. Glavin, Ajit K Roy**

*[f]Air Force Research Laboratory, Materials and Manufacturing Directorate, Wright-Patterson Air Force Base, Dayton, OH 45433, USA*

# Abstract

Miniaturization of electronic components has led to overheating, increasing power consumption and causing early circuit failures. Conventional heat dissipation methods are becoming inadequate due to limited surface area and higher short-circuit risks. This study presents a fast, low-cost, and scalable technique using 2D hexagonal boron nitride (hBN) coatings to enhance heat dissipation in commercial electronics. Inexpensive hBN layers, applied by drop casting or spray coating, boost thermal conductivity at IC surfaces from below 0.3 W/m-K to 260 W/m-K, resulting in over double the heat flux and convective heat transfer. This significantly reduces operating temperatures and power consumption, as demonstrated by a 17.4% reduction in a coated audio amplifier circuit board. Density functional theory indicates enhanced interaction between 2D hBN and packaging materials as a key factor. This approach promises substantial energy and cost savings for large-scale electronics without altering existing manufacturing processes.

# Introduction

The rapid miniaturization and integration of electronic components have resulted in enhanced heat-generating spots in electronic devices [1,2]. Microprocessors and other semiconductor devices tend to consume higher power when overheated due to excess leakage, changes in electric resistance, and adjustments needed to maintain switching efficacy [3]. Overheating is also the leading cause of failure of microprocessor chips, for example due to heat trapping in fins of fin FETs that leads to dielectric breakdown [4], and electromigration induced failure of metal interconnects [5]. Effective heat removal strategies are hence in huge demand for reducing power consumption and for extending the longevity of sensitive electronic devices. Currently, heat removal in bulk electronics is mainly done through heat sinks, liquid cooling, jet impingement and heat pipes [6]. However, miniaturizing and safely implementing these methods are reaching their limits in modern microelectronic circuit boards where a lot of components are densely packed.

One area that has received relatively less attention is developing methods for improving the thermal conductivity at the packaging-air boundary. Improving this interface with an appropriate coating would substantially enhance existing cooling approaches. In this context, diamond-based heat sinks with optimized surface coatings and SiC heat sinks using extrusion freeform fabrication (EFF) technique and nanosecond laser micro nano processing which exhibit superior thermal conductivity have been proposed for assisting in heat management [7,8]. However, their practical applications remain limited due to their high cost of fabrication and the requirement of binders that tend to be thermally resistive. Two-dimensional materials such as graphene, hexagonal boron nitride (hBN), and transition metal dichalcogenides (TMDs) are also attractive for thermal management. 2D materials exhibit ballistic and hydrodynamic phonon transport at ~100K and show thickness dependencies [9–12] suggesting superior thermal conductivities at lower dimensions. In thermal management, the application of Fourier's law of heat conduction highlights several critical factors impacting the thermal conductivity of 2D materials, including length, thickness, and engineering modifications [13]. Among 2D materials, graphene has been widely explored due to its excellent thermal conductivity of 4800-5800 W/m-K [14,15]. However, graphene's conductive nature poses a risk of short circuits in electronic devices, making it less suitable for direct application in electrically sensitive environments. In contrast, hexagonal boron nitride (hBN) combines high thermal conductivity with electrical insulation and mechanical flexibility, offering an attractive alternative for electronic cooling applications [16]. Theoretical studies based on the Boltzmann transport equation suggest that monolayer hBN exhibits thermal conductivities exceeding 600 W/m-K at room temperature, attributed to reduced phonon-phonon scattering compared to bulk hBN (~400 W/m-K) [17].

For nanoelectronic devices, hBN's compatibility with van der Waals (vdW) heterostructures, such as those formed with transition metal dichalcogenides (MX: $MoS_2$, $WSe_2$, $WS_2$), has demonstrated enhanced interfacial thermal conductance, thereby improving thermal management [18]. For bulk electronic applications, hBN has been incorporated into composites with binders such as polyvinylidene fluoride (PVDF) and other thermal interface materials to manage heat in devices like MOSFETs [19–21]. However, the low thermal conductivity of these binders reduces the composite's overall effectiveness. A binder-free coating approach, as explored in our work, could significantly improve heat dissipation by providing a more direct thermal pathway.

2D hBN can be synthesized in large quantities through scalable methods such as ball milling and liquid-phase exfoliation [22]. While ball milling is effective for industrial-scale production, liquid-phase exfoliation enables the direct formation of stable hBN dispersions, which simplifies the coating process. Selecting a suitable solvent is essential for achieving a well-dispersed, low-toxicity suspension that is compatible with commercial applications. Iso-propyl alcohol (IPA) has been identified as an ideal solvent for producing 2D hBN dispersions, achieving ~50% yield (0.06 mg/mL hBN nanosheets) with strong solvent-hBN interactions [17]. IPA's low electrical conductivity and volatility make it advantageous for coating electronic devices, as it evaporates quickly, leaving a uniform 2D hBN film. This binder-free coating approach takes advantage of the high surface area and strong Coulombic interactions of 2D hBN, which ensure adhesion to the substrate [23]. Given the multidirectional heat flow in electronics, these coatings can be tailored to address both in-plane and cross-plane thermal dissipation, providing an efficient solution for managing localized hotspots.

In this work, we show such stable, low-cost, scalable and binder-free thermal interfaces that significantly improve thermal management in electronic devices. Significantly enhanced heat dissipation was achieved in commercially-manufactured integrated circuit (IC) chips and microelectronic components coated with 2D-hBN through facile drop-casting or spray-casting techniques. These coatings were applied from 2D-hBN dispersions in isopropanol, obtained through conventional liquid phase exfoliation methods. The detailed information regarding characterization of 2D hBN is provided in the supporting information. Large commercial boards could be coated in tens of minutes without the need of any binders. We present detailed investigations of the thickness dependent thermal properties of these 2D-hBN interfaces through laser interferometry. In addition, we present density functional theory results that validate the thermal transport and interaction mechanism between 2D-hBN and the IC packaging polymer. Finally, we demonstrate real-world improvement of thermal management and performance of a commercial micro audio amplifier circuit board using our 2D-hBN spray coating.

## Results and Discussion

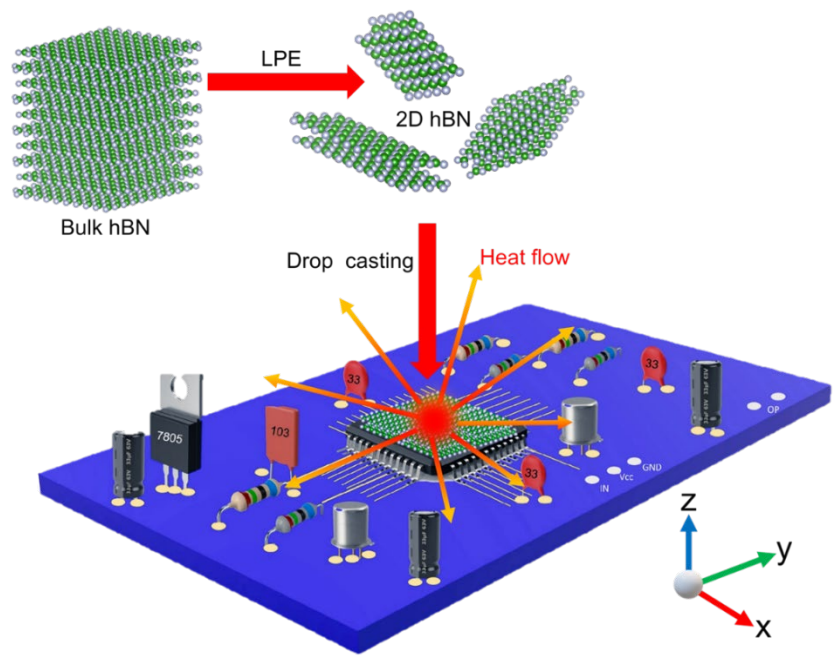

*Figure 1. Fabrication of 2D hBN coating.* Schematic representation of the synthesis of 2D hBN from bulk hBN through liquid phase exfoliation method. The obtained 2D hBN is then drop-cast onto to IC chip of an electronic device. The coating acts as a heat extraction source with enhanced heat dissipation.

**Figure 1** shows the schematic representation of the experiment. To examine the morphology and particle size distribution of the exfoliated hBN, we conducted Atomic Force Microscopy (AFM) studies. **Figure 2a** shows the AFM image of exfoliated hBN flakes with thicknesses ranging from 0.8 nm to approximately 6 nm, corresponding to monolayer to few-layer structures. These flakes are uniformly distributed across the substrate surface, with lateral dimensions ranging from approximately 100 to 150 nm, as shown in **Figure 2b**. The size distribution analysis based on the AFM image in **Figure 2a** indicates that most flakes have thicknesses in the range of 1.5 to 2 nm, as presented in **Figure 2c**. This distribution confirms the presence of a mixture of monolayer, bilayer, and few-layer hBN flakes in the sample, supporting the uniformity and reproducibility of our exfoliation and coating process. We then utilized these to study the effect of 2D hBN on heat management in electronic circuits. Initially we analysed the variation of thermal conductivity of varying thickness 2D hBN coatings spray coated on the surface of an audio amplifier integrated circuit (IC) chip. **Figure 2d** shows a photograph of the IC surface coated with 2D hBN, including an enlarged view of only the 2D coating obtained through 3D scanning of the surface. **Figure 2e** shows the cross-sectional FESEM image of the 2D hBN-IC interface with a coating thickness of ~ 200 μm. The interaction of 2D hBN coating on the IC packaging polymer - epoxy cresol novolac (ECN) - was analysed through Raman spectroscopic studies (**Figure 2f**). Several peaks are observed peaks at 1250cm$^{-1}$, 1280cm$^{-1}$ and 1596cm$^{-1}$ which belongs to vibrations of epoxide ring of ECN polymer [24]. Whereas, the $E_{2g}$ mode (in-plane phonon mode) of hBN shows a shift towards a lower wavenumber in the case of the coating indicating possible interaction with ECN polymer. The phonon modes in the 2D hBN coatings were characterized using Raman mapping, as shown in **Figure S2**. The Raman spectra for the coated IC surface reveal a prominent $E_{2g}$ phonon mode (at 1364 cm$^{-1}$), indicating an in-plane orientation of the 2D hBN. To further assess the thermal properties, we measured the in-plane thermal conductivity of the 2D hBN coatings at various thicknesses, as well as the bare IC package (ECN), which is presented in **Figure 2g**. It can be seen that the thermal conductivity of IC plastic package is significantly low at room temperature (~ 0.3W/m-K) and decreases as the temperature increases. This is because as temperature increases the spacing between adjacent polymer chains increases and order reduces through rotation in the chains. This led to an overall decrease in thermal conductivity and lowered heat spreading capability making it less effective for heat dissipation in electronic components. Whereas, in the case of 2D hBN coatings the thermal conductivity increases with coating thickness as seen in **Figure 2h**. Since phonons govern heat transport in hBN due to its ultra-wide bandgap, high crystallinity, low atomic mass, and elevated Debye temperature, these properties collectively lead to a high phonon group velocity, extended phonon mean free path, and high vibrational frequencies, all of which reduce phonon scattering and contribute to enhanced thermal conductivity, particularly in the in-plane direction [17,25,26]. Additionally, the strong mechanical integrity of hBN supports effective phonon transport [27]. Our spray coating technique allows for controlled increases in coating thickness, where greater thickness prolongs atomic vibrations, further facilitating phononic transport.

Finally, **Figure 2i** shows the thermal conductivity of other materials typically used for thermal management [28–34]. A majority of these materials are composites consisting of polymer binders and highly thermal conductive materials like graphene. However, due to the addition of binders the overall thermal conductivity is significantly low. In contrast, the drop casted 2D hBN coating on ECN exhibits superior thermal conductivity suggesting superior heat transfer capabilities of 2D hBN coatings.

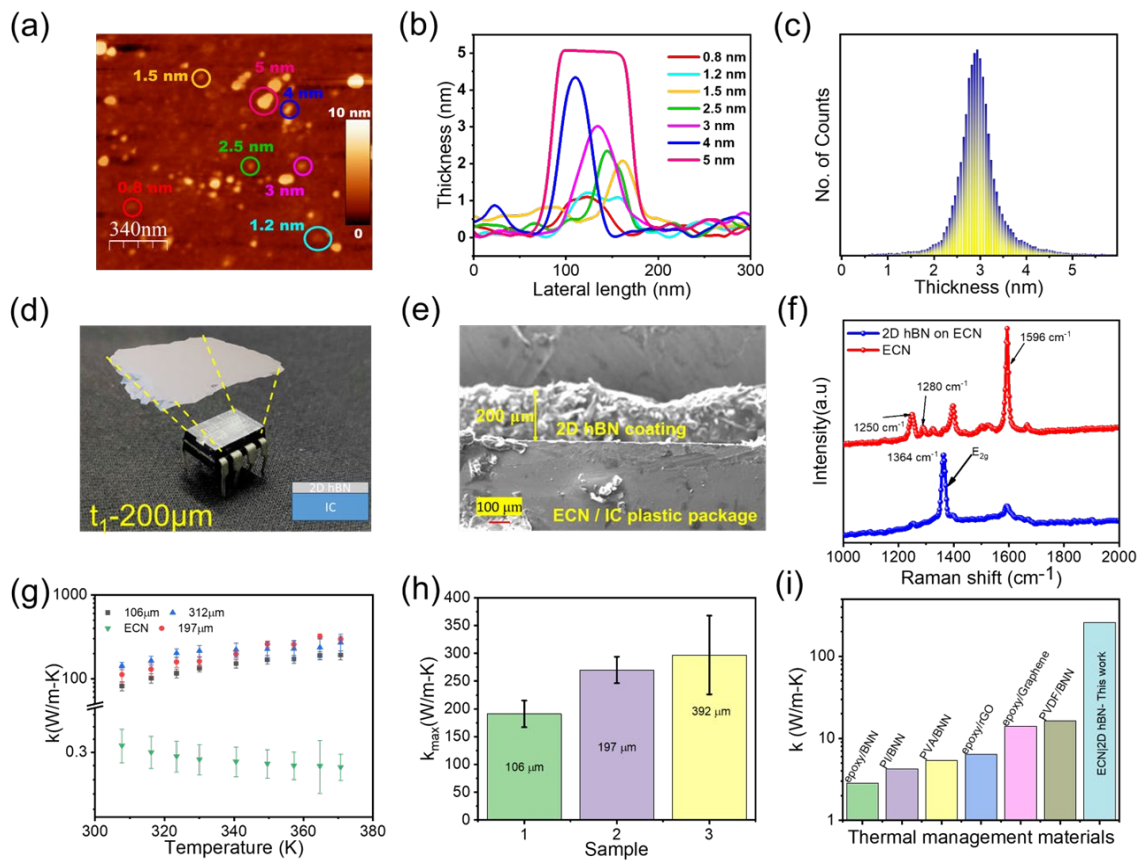

*Figure 2. Thermal conductivity, Raman spectroscopy of 2D hBN coatings*. **a**, AFM image of exfoliated hBN showing particles of various thickness. **b**, Thickness and lateral length profiles of exfoliated hBN. **c**, Size distribution plot of exfoliated hBN. **d**, Photograph of 2D hBN coating on Integrated circuit (IC) polymer package (ECN). **e**, FESEM image of the cross-section of 2D hBN coating-IC plastic package showing the coating thickness. **f**, Raman spectra of 2D hBN coating on the plastic package of a tested IC. **g**, In-plane thermal conductivity plots of bare IC surface and 2D hBN coating. **h**, Maximum In-plane thermal conductivity plots with varying thickness coatings. **i**, Thermal conductivity of other reported thermal management materials.

To determine the rate of heat flux and convective heat transfer coefficient of 2D hBN coatings we performed Mach-Zehnder interferometric (MZI) studies. The convective heat transfer coefficient and heat flux for coated and uncoated IC chips were investigated using MZI adopting Naylor's approach [35,36]. **Figure 3a** shows the experimental setup for interferometry studies and wide field fringe diagram with fringe angle (α). The fringe diagram shows an IC surface where we make the coating. The fringe angle indicates the magnitude of heat flux from the IC surface. By measuring the fringe angle with respect to IC surface we can calculate the heat flux and heat transfer coefficient. The detailed information of the calculation of heat transfer coefficient and heat flux measurements using fringe angle is mentioned in methods section. **Figure 3(b-e)** shows the 3D image of fringes obtained in the case of initial, non-coated, bulk hBN-coated, and 2D hBN-coated IC chips, respectively. The interferograms show the highest fringe angle deflection in the case of the IC chip coated with 2D hBN, followed by the chip with bulk hBN coating and the chip with no coating. **Figure 3f** and **Figure 3g** show the comparison of the heat flux and heat transfer coefficient respectively, for the bare and 2D hBN coated IC chips. The maximum heat transfer coefficient was observed in the case of the IC chip coated with 2D hBN, followed by that coated with bulk hBN. A similar trend was observed for the heat flux also with maximum heat flux obtained in the case of an IC chip coated with 2D hBN. We find the remarkable result that simply coating a commercial IC chip with 2D hBN can more than double both the heat transfer coefficient as well as the heat flux. The higher heat transfer coefficient and heat flux value observed in the case of an IC chip coated with 2D hBN could be attributed to the enhanced thermal conductivity of the coating as observed in thermal conductivity studies and the increase in surface area. Thus, the application of a 2D hBN coating on the surface of the IC chip is a simple and highly effective method for enhancing the heat dissipation from the IC chips.

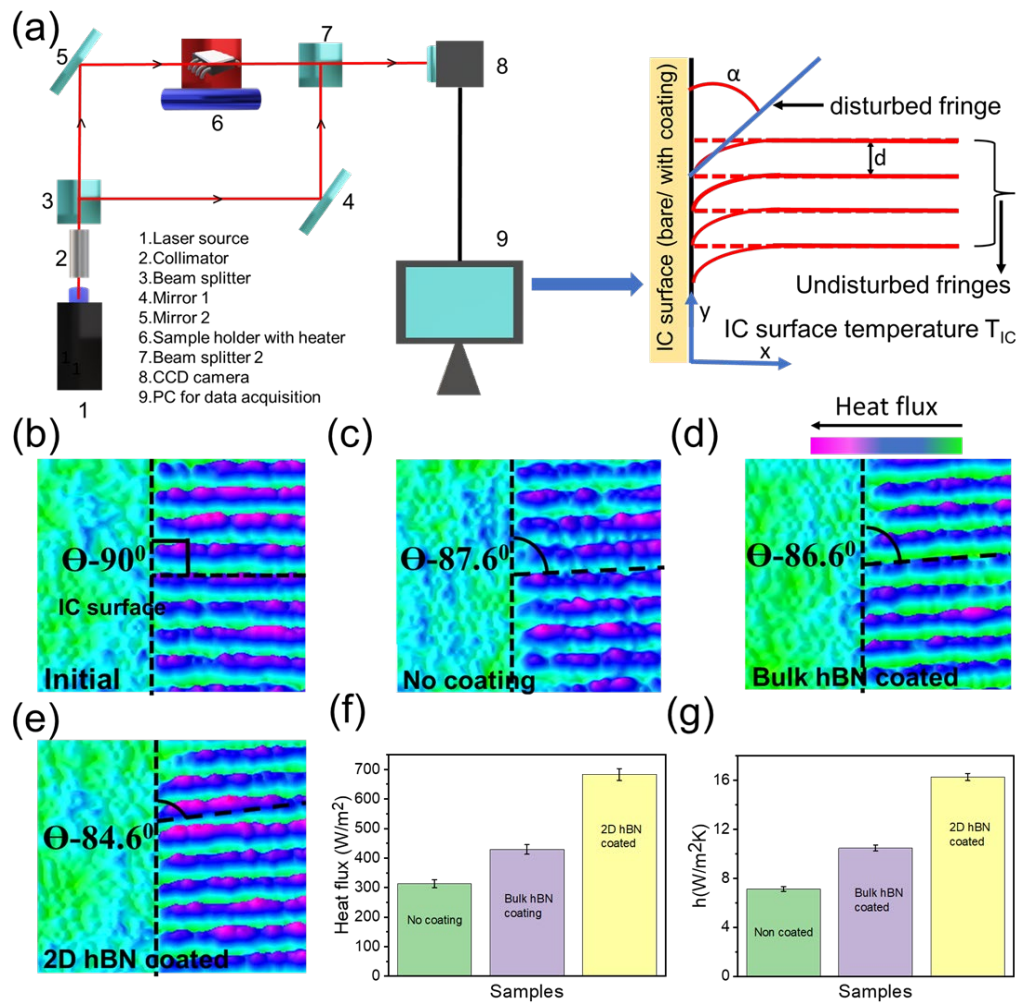

*Figure 3. Interferometric studies on 2D hBN coatings*. **a**, *Schematic representation of the experimental setup and wedge fringe field formation adjacent to a hot surface ray diagram*, **b**, *Interferograms of initial.* **c**, *non-coated.* **d**, *bulk hBN.* **e**, *2D hBN coated IC showing variation in fringe angle deflection.* **f**, *Heat flux profiles of non-coated, bulk hBN coated and 2D hBN coated samples.* **g**, *Convective heat transfer coefficient plots of non-coated, bulk hBN coated and 2D hBN coated samples.*

To demonstrate the in-operando heat dissipation improvement in 2D hBN-coated electronics, we drop casted 2D hBN onto the entire surface of an audio amplifier circuit board and performed thermal imaging studies. **Figure 4a** shows the experimental setup for thermal imaging consisting of a thermal imaging camera and a PC for data acquisition. The 2D hBN coated circuit board is shown in **Figure 4b**. **Figure 4c** compares the thermal image of bare and 2D hBN coated circuit boards. This comparison provides clear visual evidence of heat trapped at main IC and other electronic components in the case of bare circuit board, whereas in the 2D hBN coated circuit board, heat is more effectively dissipated through the coatings. **Figure 4d** shows the variation of surface temperature from identical spots of the circuit board (marked 3 in **Figure 4c**) with and without coating, as the board is first turned ON and then OFF after 140 s. The hBN-coated circuit board shows clear higher temperature trend along with heat branching from the main IC , reaching a difference of 1.2 °C compared to the bare circuit board. This is due to heat flux dissipation through 2D hBN channels as observed in the interferometric studies as a result of high thermal conductivity of 2D hBN coatings. **Figure 4e** shows the maximum temperature differences between coated and uncoated circuits, from other electronic components (labelled 1-7 in **Figure 4c**) which confirm the circuit-board-wide improved removal of heat in the 2D hBN-coated. Finally, to assess the power consumption we measured the time dependent current and calculated power measurements as shown in **Figure 4f** and **Figure 4g**. The power was calculated using the relation P= VI, where P is the power consumed, V is the operating voltage (9V in our case) and I is the current consumed. **Figure 4h** compares the maximum power consumption in both (bare *vs*. 2D hBN coated device) cases where a 17.4% reduction in power consumption is observed for 2D hBN coated audio amplifier. This is due to the fact that as per Joule's law heat generated in an electrical device is proportional to square of current and lowering this current by means of any cooling methods could significantly lower current and there by overall power consumption. This finding shows the effectiveness of 2D hBN coatings in lowering the power consumed.

.

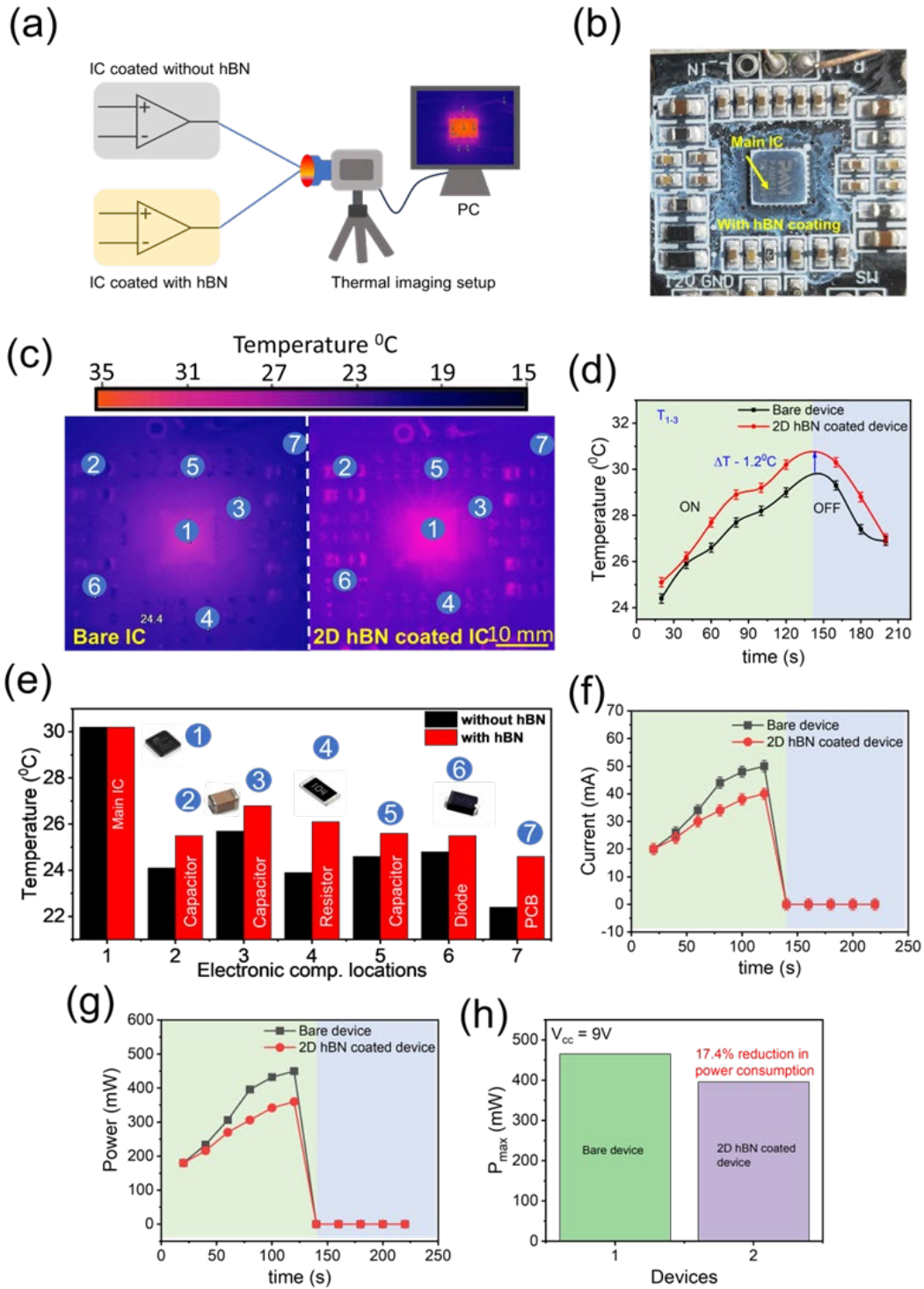

*Figure 4. Thermal imaging studies of heat dissipation of audio amplifier circuit board to study the enhanced heat dissipation in 2D hBN coated electronics. **a**, Schematic representation of the experiment, **b**, Photograph of audio amplifier circuit board with 2D hBN coating. **c**, Thermal image of bare and 2D hBN coated circuit board, the numbers from 1-7 indicates various electronic components on the circuit board. **d**, Time-dependent heating-cooling profile of region marked as 3 from the main IC 1 of the audio amplifier. **e**, Temperature profiles of various components on the audio amplifier. **f**, Time dependent current consumption of the bare and 2D hBN coated amplifier circuit board. **g**, Time dependent power consumption*

*of the bare and 2D hBN coated amplifier circuit board. h, Comparison plot of maximum power consumed by two devices.*

With the experimental confirmation of enhanced heat dissipation capability of 2D hBN coating, we further performed theoretical studies to understand the thermal transport mechanisms in 2D hBN using first-principles calculations. To illustrate the experimentally observed enhancement in the in-plane $\kappa_l$ values with the increase in the coating layers (along the z-axis), we analyzed bonding and phononic transport parameters of monolayer, bilayer, and bulk. The optimized lattice parameters of monolayer, bilayer, and bulk are given in **Table 1** in agreement with the reported values [37,38].

*Table 1: Lattice parameters of monolayer, bilayer, and bulk hBN. $I_{sep}$ is the interlayer separation among the layers*

| Structure | a, b (˚A) | c (˚A) | $I_{sep}$ (˚A) |
|---|---|---|---|
| Monolayer | 2.512 | 19.533 | - |
| Bilayer | 2.505 | 19.645 | 3.297 |
| Bulk | 2.503 | 6.441 | 3.220 |

We have also found that AA´ stacking is energetically more favourable than AA stacking for the bilayer and bulk [39]. The aforementioned AA stacking was further used to calculate thermal transport properties. The binding nature has been analyzed using the electron localization function (ELF) [40]. ELF has been assigned the values ranging from 0-1, where one means perfect localization corresponding to covalent bonding (red colour) and zero to absolute delocalization means no bonding (blue colour) among atoms. The calculated ELF values (in yz-plane) have been plotted in **Figure 5a** corresponding to monolayer, bilayer, and bulk, respectively. A significant difference in the ELF values along the in-plane (y-direction) can be seen compared to the out-of-plane(z-direction). All three structures show ELF=0 along the z-directions, suggesting no bonding along that axis among atoms/layers. However, the ELF value increases among atoms with the thickness of the structure (monolayer to bulk) along the y-direction (see **Figure 5a**). Thereby suggesting increased in-plane bonding strength between atoms with the structure's thickness. Stronger bonding leads to strong atomic vibrations, which can inherently lead to an enhanced heat-carrying ability of the system under consideration [41], aligning with our experimental observations.

To comprehend our ELF analysis, we calculated $\kappa_l$ as a function of phonon frequencies at 400K. The calculated in-plane $\kappa_l$ values have been shown by black lines in **Figure 5b** for monolayer, bilayer, and bulk, respectively. The variation in lattice thermal conductivity ($\kappa l$) with frequency is highlighted by red-filled regions in **Figure 5b**. As observed, $\kappa l$ increases with an increase in structural thickness along the z-direction, attributed to enhanced in-plane bonding strength. While direct comparison of these DFT-calculated $\kappa l$ values with experimental results is not feasible—since the DFT calculations are performed at nanoscale thicknesses (~$10^{-10}$ meters) whereas experiments occur at microscale thicknesses (~$10^{-8}$ meters)—the trend of in-plane $\kappa l$ enhancement with sample thickness along the z-direction is consistently observed in both experimental and theoretical analyses. Notably, phonon frequencies around ~800 cm$^{-1}$ predominantly contribute to $\kappa l$, underscoring the role of these frequencies in governing thermal transport properties.

To further explore the phononic behavior of hBN with varying thicknesses, we analyzed key thermal transport parameters: phonon group velocity ($v_g$), phonon lifetime ($\tau$), and phonon mean free path. **Figure 5c** illustrates these parameters as a function of cumulative $\kappa l$. While $v_g$ remains relatively constant across different thicknesses, $\tau$ and phonon mean free paths significantly increase with thickness. Specifically, the rise in $\tau$ from monolayer to bulk (**Figure 5d**) suggests that greater thickness extends atomic vibration duration, enhancing phononic heat transport. This increase in $\tau$ is directly correlated with an extended phonon mean free path, as shown in **Figure 5e**. Overall, our computational results demonstrate that increased thickness along the z-direction in hBN strengthens in-plane bonding, concurrently boosting phonon lifetime and mean free path, thereby resulting in enhanced in-plane κl values.

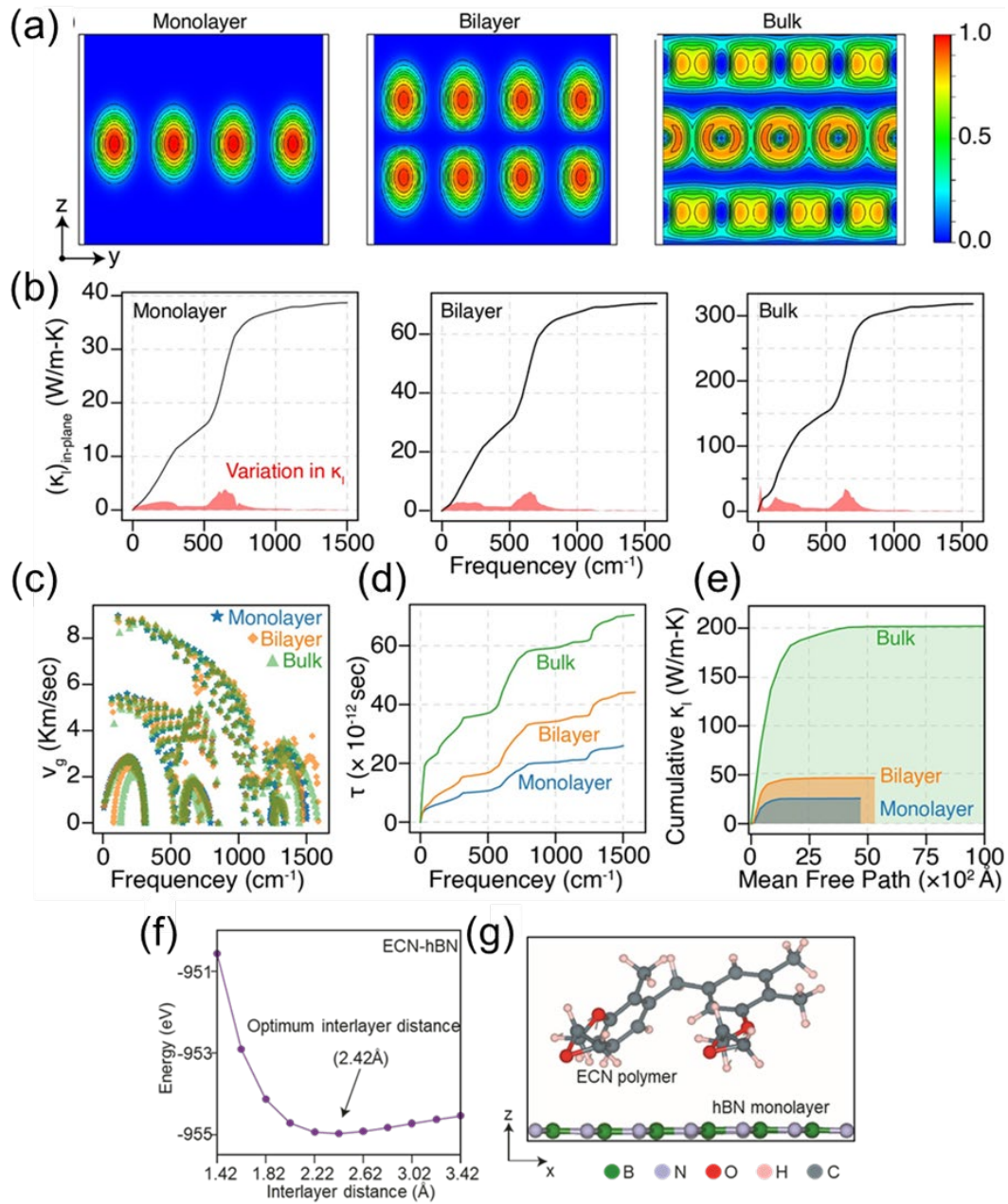

***Figure 5.*** *Computational thermal transport and interaction studies in 2D hBN-ECN. **a**, Electron localization function of the monolayer, bilayer and bulk in yz-plane. **b**, In-plane lattice thermal conductivity (κ$_l$) of the monolayer, bilayer, and bulk, respectively at 400K. **c**, phonon group velocity (v$_g$) as a function of phonon frequency of monolayer, bilayer, and bulk represented by asterisk, diamond, and triangle, respectively. **d**, phonon lifetime (τ) as a function*

*of phonon frequency of monolayer, bilayer, and bulk represented by blue, orange, and green, respectively at 400K. **e**, phonon mean free path as a function of cumulative $\kappa_l$ of the monolayer, bilayer, and bulk represented by blue, orange, and green fill, respectively at 400K. **f**, Optimization of the interlayer distance between ECN-polymer and hBN-monolayer. **g**, ECN-hBN-monolayer fully optimized structure.*

Next, we show the interactions between ECN polymer and hBN-layers. To mimic the ECN polymer, a 3D structure with two connected ECN monomers is chosen [42]. The optimized structure of the modelled ECN is then placed over the already optimized hBN monolayer, bilayer, and tetralayer systems to analyze the stability of ECN-hBN heterostructure having hBN with different thicknesses. First, the interlayer distance of the heterostructure (ECN-hBN-monolayer) was optimized as shown in **Figure 5f**, and the entire structure was optimized thereafter. We found that the interlayer distance between ECN-hBN-monolayer is less than ~1°A than the hBN-hBN interlayer distance, as given in **Table 1**. Thereby, suggesting the possibility of interaction between ECN and hBN layers. The optimized structure of the ECN-hBN monolayer is shown in **Figure 5g**. The interaction between the ECN and hBN is evaluated based on the formation energies, a thermodynamic stability index, of these heterostructure ECN-hBN systems. The formation energy ($E_f$) is calculated using the following equation:

$$E_f = E_{hetero} - E_{hBN} - E_{ECN} \qquad (1)$$

where the $E_{hetero}$, $E_{hBN}$, and $E_{ECN}$ are the total electronic energies of the ECN-hBN heterostructure, the hBN system, and the modelled ECN system, respectively. The heterostructure formation energy of the ECN and the hBN-monolayer (-0.03 eV/atom) is less than that of the ECN and the hBN-bilayer (-0.01 eV/atom) and the ECN and the hBN-tetralayer (-0.01 eV/atom), suggesting that the ECN will strongly interact with the hBN-monolayer compared to the multilayer hBN. Thereby, with an increase in hBN-thickness, the stability of the ECN-hBN heterostructure reduces.

Our experimental and DFT studies demonstrate that spray-coated 2D hBN coatings provide an effective solution for heat management in both bulk and microelectronic devices. Importantly, the direct synthesis of 2D hBN dispersion through liquid-phase exfoliation eliminates the need for additional solvents in the coating process, reducing production costs. This method also allows precise control over key parameters, such as coating thickness, area coverage, and application time. Traditional heat management approaches—such as metallic heat sinks and liquid cooling—significantly increase device costs. Additionally, liquid cooling requires complex pumping systems, limiting its use to bulk electronics, and its dense components add to device weight. In contrast, 2D hBN coatings address these limitations, offering a lightweight and efficient alternative without these drawbacks. Notably, heat sinks are challenging to integrate with microelectronic components due to limited contact areas and potential risks of short-circuiting. In such cases, ultra-thin 2D hBN coatings, with their high thermal conductivity and low electrical conductivity, provide an ideal solution for heat dissipation without compromising performance. We propose that applying spray-coated 2D hBN across entire circuit boards could act as an effective heat sink, ultimately lowering power consumption and enhancing device longevity.

## Conclusions

The current work demonstrates enhanced heat dissipation in electronic chips using binder-less 2D hBN coatings. 2D hBN was synthesised using liquid phase exfoliation of bulk hBN. Later the coating was made by drop casting 2D hBN containing dispersion onto the surface of electronic components. Thermal conductivity measurement showed thickness dependencies which was justified by heat flux measurements using interferometric studies. Thermal imaging studies on a 2D hBN-coated audio amplifier circuit board revealed an enhanced heat dissipation leading to low power consumption. The above thermal transport findings were also validated using DFT studies which also verified the interaction between 2D hBN and ECN, suggesting the unnecessity of using any binder. The cost of industrial grade hBN used was less than 1 cent to coat a full PC motherboard. Our findings suggest that binder-less 2D hBN coatings as promising materials for efficient heat dissipation in electronics.

## Methods

### Synthesis 2D hBN coating

2D hBN was synthesised by liquid phase exfoliation of 5g bulk hBN (97%, Sigma-Aldrich) powder in Iso-Propyl alcohol (Sigma-Aldrich purity ≥ 99%) solvent (50 mL) for 2 hours. The obtained dispersion was centrifuged for 1 hour to separate 2D hBN sheets. To make the coating the 2D hBN containing dispersion was drop cast onto to Integrated IC chip (TDA2822) surface at $50^0$C for 15-30 minutes until a thin white coating was observed. The thickness of the coating was varied by varying the time of drop casting. The thermal conductivity of drop casted 2D hexagonal boron nitride (hBN) is measured using a Physical Property Measurement System (PPMS) with a microcalorimeter. The sample is exposed to a thermal pulse in a high vacuum, and its temperature response is recorded to determine its heat capacity. The software allows for measurements of heat capacity with automated field calibration.

### Thermal Transport Measurement

The 2D hBN was spray-coated onto the polymer substrate and then carefully mounted on the PPMS thermal transport puck, which is optimized for thin samples. A suitable adhesive or thermal grease was applied to ensure good thermal contact between the sample and sensors without influencing the intrinsic thermal properties. The PPMS thermal conductivity module employs two thermometers and a heater, with the heater attached to one end of the sample and thermometers positioned at both ends. To perform the measurement, the PPMS was set to apply a controlled heat pulse via the heater, generating a steady-state temperature gradient across the sample. By monitoring the temperature difference between the two thermometers and

recording the power input from the heater, we calculated the thermal conductivity (k) using Fourier's law:

$$k = \frac{Q.L}{A.\Delta T}$$

where k represents the thermal conductivity, Q is the heat flow, L is the distance between the thermometers, A is the cross-sectional area of the sample, and ΔT is the temperature difference.

We measured the thermal conductivity of both the bare polymer and the hBN-coated polymer to highlight the significant contrast between the low thermal conductivity of the polymer and the enhanced values due to the hBN coating. The heat pulse was carefully controlled to avoid excessive heating that could impact the measurements. Additionally, for coatings of varying thickness, the thermal conductivity was determined by accounting for layer effects. Given the high in-plane thermal conductivity of hBN, corrections or approximations were applied to decouple the thermal contributions of hBN and the polymer. To ensure accuracy, we repeated the measurements at different temperatures and heat fluxes to assess stability and evaluate any temperature dependence in the thermal conductivity of the hBN-polymer composite.

**Mach-Zender Interferometry**

Mach-Zehnder Interferometry (MZI) was used in the present work to estimate the heat dissipation from the IC chips coated with bulk and 2D hBN. MZI utilizes the interference of two coherent monochromatic light beams to estimate the local variation in refractive index, density, and temperature of the medium. The interference fringe pattern gets distorted when an additional phase delay is introduced in the path of either of the beams, caused by changes in the refractive index of the medium due to temperature-induced variations in density. The heat transfer parameters were estimated from the distortion of fringe patterns and thus MZI is a straightforward and non-intrusive method for quantifying the heat dissipation. MZI setup consists of a diode laser (wavelength of 650 nm, 2 mW), as the source of monochromatic light. The optical configuration comprises a beam expander or collimator to enlarge the beam to cover the entire field, beam splitters (half-silvered mirrors) to divide and recombine the light beams, front-coated mirrors for guiding the beams, and a focusing lens to converge the beams onto the CCD camera. Wedge fringe mode, in which the fringes are perpendicular to the heat transfer surface, was adopted in the present work. The distorted fringes are captured by a CCD camera and analysed using image processing software to estimate the heat transfer parameters. The experimental setup is provided is the supporting information (**Figure S3**).

The heat transfer coefficient was estimated from the fringe distortion angle using Naylor's methods [43], as follows:

$$h = \frac{2\lambda R K_a T_s^2}{3\bar{r} d Z P (T_s - T_a) \tan\alpha} = \frac{C}{\tan\alpha} \qquad (2)$$

The various parameters in the equation are,

λ = Laser beam wavelength (650 nm),

R = Universal gas constant (8.314 J/kgK),

Ka = Thermal conductivity (W/m-K),

Ts = Surface temperature (K),

r̄ = Specific refractivity of air,

d = Fringe width (m),

α = angle between line of constant fringe shift and surface (°)

Z = Length of the test piece in the direction of the laser beam (m),

P = Ambient pressure (Pa),

Ta = Ambient temperature(K),

G = Gladstone - Dale constant (m$^3$/kg)

The numerical value of the specific refractive index of air ($\dot{r}$) is $1.504 \times 10^{-4}$ $m^3/kg$. The Gladstone-Dale constant ($G$) for air is $2.24 \times 10^{-4}$ $m^3/kg$.

The local heat flux was estimated from the heat transfer coefficient as follows.

$$q(t) = h(T_s(t) - T_a) \qquad (3)$$

The heat transfer studies were conducted on the chip in the MZI set up and the heat transfer parameters were estimated. Optical components were initially aligned to obtain parallel fringes, which are perpendicular to the surface of the chip. The IC chip was heated using a flat plate heater (5 W), keeping the legs of the chip in contact with the heater plate surface. The heater plate was properly insulated on all sides to prevent heat loss to the surroundings. The heater plate block along with the IC chip was placed in the path of one of the beams of the interferometer and heated. Once the heater is ON, the IC chip surface begins to dissipate the heat, causing the fringes to bend proportionally. The fringes were captured using a CCD camera once the surface of the chip reached $70^0$C. The temperature of the IC chip was measured using an Infrared thermometer and the ambient temperature was measured using a K-type thermocouple and was recorded by data acquisition system (Agilent 34972A). The fringes obtained were analyzed using Motic Images Plus 2 software to obtain the fringe deflection angle and fringe width. The convective heat transfer coefficient (h) and the heat flux (q) for the IC chip were determined using **Eq. 2, 3**, respectively, as previously stated.

## Computational Details

First-principles density functional theory (DFT) calculations have been performed using the Vienna *Ab-initio* Simulation Package (VASP) [44,45]. Electron-ion interactions have been considered by using projected augmented wave [46,47] pseudopotentials. Electronic exchange and correlations have been incorporated by utilizing Predew-Burke-Ernzerhof (PBE) [48] functional within the generalized gradient approximation (GGA). The conjugate gradient approximation optimized the monolayer, bilayer, and bulk until the Hellmann-Feynman forces operating on each atom were smaller than 0.00005 eV/˚A. The kinetic energy cutoff was set to 400 eV. A dense Γ-centered Monkhorst-Pack [49] k-grid of 12×12×1 (for monolayer and bilayer) and 12×12×6 (for bulk) is used to sample the Brillouin zone (BZ). In the case of monolayer and bilayer, a vacuum of 16 ˚A was employed to avoid any spurious interaction between the periodic images. Van der Waals interactions have been considered by applying the DFT-D3 method of Grimme [50]. The harmonic interatomic force constants (IFCs) have been calculated considering 8×8×1 (for monolayer) and 5×5×1 (for bilayer and bulk) supercell. A strict electronic energy convergence of $10^{-8}$ eV has been used. The finite displacement method has been utilized to obtain phonon frequencies implemented in the Phonopy [51]. Anharmonic third-order IFCs have been calculated using 7×7×1, 4×4×1, and 5×5×1 supercell sizes for monolayer, bilayer, and bulk, respectively. Utilizing the obtained harmonic and anharmonic IFCs, lattice thermal conductivity ($\kappa_l$) has been calculated by solving the phonon Boltzmann transport equation (PBTE) under relaxation time approximation [52]. Long-range interactions were considered with Born effective charges and calculated using the density functional perturbation theory.